# The Role of Protein Electrostatics in Facilitating the Catalysis of DEAD-box Proteins

**Christopher M. Frenz**[1]
[1]Department of Computer Engineering Technology
New York City College of Technology (CUNY)
Brooklyn, NY 11201
cfrenz@citytech.cuny.edu

**Abstract -** *Protein electrostatic states have been demonstrated to play crucial roles in catalysis, ligand binding, protein stability, and in the modulation of allosteric effects. Electrostatics states are demonstrated to appear conserved among DEAD-box motifs and evidence is presented that the structural changes that occur to DEAD box proteins upon ligand binding alter the DEAD-box motif electrostatics in a way the facilitates the catalytic role of the DEAD-box glutatmate.*

## 1 Introduction

Over the course of the last decade, DEAD-box proteins have arisen as an important class of proteins that are involved in a wide range of activities related to RNA synthesis and function. The primary functions of most DEAD-box proteins are believed to be ATP-dependant RNA helicases, and have been characterized as having roles in activities such as mRNA preprocessing, translation, ribosome synthesis, and RNA complex formation [1].

DEAD-box proteins get their characteristic name from the amino acid composition (DEAD) that comprises the protein class' motif II, one of several highly conserved motifs within DEAD-box proteins. The residues that comprise motif II have been demonstrated to play a role in the coordination of ATP in the active site via Mg2+ ion that interacts with the motifs initial Aspartate residue. The motif has also been demonstrated to provide the ability to catalyze ATP with the Glutamate residue acting as the catalytic base. Moreover, the residues that comprise motif II have been demonstrated to interact with the conserved residues of motifs I and III and these interaction patterns have been demonstrated to change under conditions where NTP binding was present or absent [2].

Recent studies on the hepatitis C virus (HCV) NS3 helicase, which is a member of the related DExH protein family, have demonstrated that electrostatic perturbations exist, as indicated by perturbed pKa values, within the DExH motif, along with numerous other residues, and have suggested that these residue perturbations may be of functional significance [3]. The functional significance of perturbed residues is further supported by studies on HCV and HIV replicative proteins that demonstrate that conserved regions exhibit a greater number and magnitude of pKa perturbations than lesser conserved regions [4] as well as by studies on other classes of enzymes that demonstrate that residues with perturbed pKas can often readily contribute to acid-base catalyzed reactions. This is especially true as the pKas of the donors and acceptors approach physiological pH, since these perturbed pKa values allow for proton transfer to occur more readily [5].

This study seeks to better define the role of the relationship between electrostatics and catalytic functionality within motif II DEAD box protein residues, analyzed from a diversity of DEAD box protein crystal structures. The study further demonstrates that electrostatic mechanisms, which contribute to the motifs catalytic ability, appear to be conserved across all DEAD box proteins studied.

## 2 Methods

### 2.1 Structure Selection

DEAD-box protein crystal structures were obtained through a search of the Protein Data Bank. The DEAD box proteins crystal structures 1HV8, 1QDE, 1QVA, 1VEC, and 2G9N structures were used to compute the pKa's of DEAD-box active sites in non-catalytic conformations [6-9]. Crystal structures 2DB3 and 2HYI each contain the bound ligand ANP as well as bound nucleotide sequences and were used to compute the pKas of precatalytic DEAD-box sites [10,11]. In cases where the PDB files consisted of multiple subunits, the pKas of all subunits were computed. ANP bound structure pKas were computed with the ligand excluded from the calculation since the H++ calculation methodology does not have good support for including ligands within electrostatic calculations.

### 2.2 Mutations

The Swiss PDB viewer was used to mutate motif 1 Lysine residues to Alanine in order to demonstrate the electrostatic contributions the motif 1 Lysines exerted on the corresponding DEAD-box residues. Alanine mutations were created for each DEAD box containing subunit of 2DB3 and 2HYI and the pKas of the residues that comprise each of these Lysine mutant proteins were calculated. Alanine mutations were also made in residues D399, R551 and H575 of 2DB3. Additional mutations were not processed due to the computationally intensive nature of the pKa calculations.

### 2.3 pKa Calculations

pKa values were calculated using the H++ Web Server, which is available at http://biophysics.cs.vt.edu/H++ [12]. This Web server begins the pKa prediction by adding missing atoms and assigning partial charges to the uploaded protien structure using the PROTONATE and LEAP modules of the AMBER molecular modelling system and the parm99 force field [13]. The positions of these added protons are then optimized using 100 steps of conjugate gradient descent minimization and 500 steps of Molecular Dynamics simulation at 300K. Protein electrostatic calculations are then performed using the Poisson-Boltzman equation in the program package MEAD, and used to compute the free energies of the protonation microstates [14]. Titration curves and pKa values are then determined using the clustering approach described by Gilson [15].

### 2.4 Figures and Plots

Protein structure representations and alignments were prepared using the Swiss PDB viewer [16] and output for rendering in POV-ray 3.5. All plots and their corresponding statistical analyses were prepared using Graph-Pad Prism 4.01.

## 3 Results

### 3.1 pKa Calculations

pKa calculations yielded conformation dependant results for amino acid residues comprising the DEAD box motif. For structures that were present in a non-catalytic state, pKa values averaged D=5.89, E=4.39, and D=2.30, whereas the formation of the pre-catalytic complex demonstrated a shift in DEAD box pKa values with averages of D=4.29, E=5.88, and D=5.01 (Figure 1). Student *t*-tests demonstrated each of the residue pKa values to be statistically different in the non-catalytic and pre-catalytic states with each *t*-test returning a *p*-value less than 0.05. These findings are suggestive that the electrostatic environment of the DEAD box residues is different in the non-catalytic and pre-catalytic states, but the similarities within each state suggest that the roles of electrostatics within the DEAD-boxes may be conserved. Moreover, the increase in the pKa of glutamate suggests that the electrostatic change supports ATP hydrolysis via glutamate, since the pKa perturbation will facilitate the role of glutamate as a catalytic base.

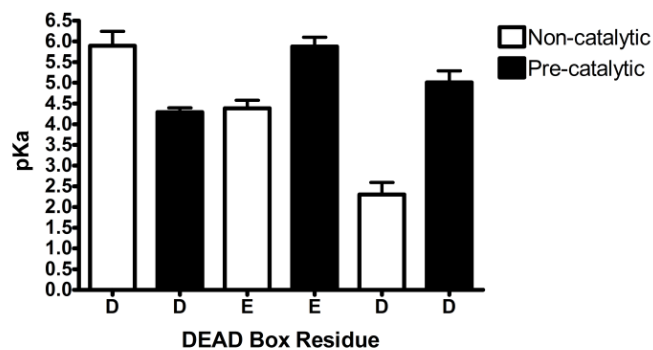

**Figure 1**- comparisons of DEAD box residue pKas in both the non-catalytic and pre-catalytic state.

## 3.2 DEAD-box Structural Comparisons

Since the presence of bound ANP ligand could not provide a direct explanation for the electrostatic changes in the DEAD-box, an examination of the residues that make strong interactions with the glutamate of the DEAD box indicates that the two charged residues that consistently interacted with the glutamate were the first aspartate residue of the DEAD-box (i.e DE) and the motif 1 lysine. Structural overlays of the various proteins revealed that in the pre-catalytic state the lysine residue is shifted away from the residues of the DEAD box. The position of the aspartate residue in relation to the glutamate remains relatively unchanged in the both the pre-catalytic and non-catalytic states structures.

## 3.3 Mutant Analysis

pKa calculations were performed on motif 1 lysine to alanine mutations of the 2DB3 and 2HYI subunits. These mutants had an average DEAD-box glutamate pKa of 4.2, which is statistically different from the unmutated pre-catalytic DEAD-box value of 5.88 (p=0.0003), but not statistically different from the non-catalytic state of the enzyme (Figure 2).

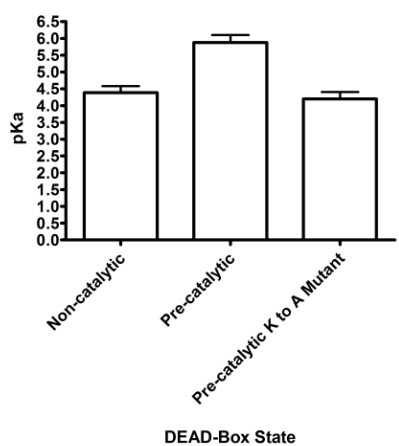

**Figure 2**- comparison of the pKa of the catalytic glutamate in the non-catalytic state and pre-catalytic state, as well as with the motif I lysine mutated to alanine.

These findings thereby suggest importance of the lysine residues' contribution to pre-catalytic state DEAD-box electrostatics. To further ascertain the role of the lysine residue, the position of the motif 1 lysine of pre-catalytic 2DB3 was shifted into the position of the motif 1 lysine of non-catalytic 1QDE and this resulted in a drop of the DEAD-box glutamate to a pKa of 3.9. Shifting the Motif 1 Lysine of 1QDE and 1HV8 into the position present in structure 2DB3 was not able to raise the pKa of the DEAD-box glutamates of these proteins to their pre-catalytic values. This suggested that while the positioning of motif I residues in relationship to the DEAD-box residues is critical to achieving a pre-catalytic electrostatic state, it is not the only residue that contributes significantly to this perturbation. A comparison between the residues having side chain interactions with the DEAD-box glutamate of non-catalytic and pre-catalytic enzymes reveals that while the motif 1 lysine and the first DEAD-box aspartate are the primary interacting residues in the non-catalytic state, there are numerous other side chain interactions present in the pre-catalytic state (Table 1). Mutational analysis of interacting residues D399, R551, H575 of the 2DB3 structure demonstrates that the contribution of each of these residues is important for the maintenance of the pre-catalytic electrostatic environment, since mutation to alanine moves the pKa of the DEAD-box glutamate away from the physiological pH range and into the range of non-catalytic state glutamate pKas (D339A DEAD-box glutamate pKa=4.2, R551A DEAD-box glutamate pKa=3.9 and H575A DEAD-box glutamate pKa=2.2). These findings thereby suggest the importance of all interacting residues being in the correct positioning in order to enable the proper electrostatic environment for catalysis.

**Table 1**: Residue Interactions with Catalytic Glutamate

| Non-Catalytic | | Pre-Catalytic | | | |
|---|---|---|---|---|---|
| Structure | Interaction | Structure | Interaction | Structure | Interaction |
| 1QDE | K88 | 2DB3 | C288 | 2HYI | K88 |
|  | D169 |  | K295 |  | E117 |
|  | E173 |  | R328 |  | D187 |
| 1HV8 | R52 |  | E329 |  | D190 |
|  | K57 |  | E337 |  | E191 |
|  | D154 |  | D399 |  | D335 |
|  | D157 |  | D402 |  | R339 |
|  | E158 |  | R403 |  | D342 |
| 1QVA | K71 |  | C462 |  | E359 |
|  | D169 |  | R551 |  | H363 |
|  | E173 |  | D554 |  | R364 |
| 1VEC | K135 |  | D571 |  | R367 |
|  | Y139 |  | D572 |  | Y371 |
|  | D235 |  | Y573 |  |  |
|  | K239 |  | H575 |  |  |
|  | Y265 |  | R576 |  |  |
| 2GN9 | D182 |  | R579 |  |  |
|  | E186 |  |  |  |  |

# 4   Discussion

The results of this study demonstrate that within DEAD box proteins, the electrostatic environment of the catalytic site in both the non-catalytic and the pre-catalytic state are conserved in several distinct DEAD-box proteins, as indicated by the similarities present between the pKa values of the residues that comprise the DEAD-box.  Recent studies have suggested an interrelationship between protein electrostatics and residue conservation, with conserved residues in HIV and HCV replicative proteins exhibiting a greater number and magnitude of pKa perturbations than non-conserved residues [4].  Among the proteins analyzed in this study was the HCV NS3 helicase which bears a related DExH motif, and is evolutionarily related to DEAD-box family proteins, which is suggestive that this relationship might also hold true for the proteins examined in this study.  The idea of conserved protein electrostatic being present among related proteins is further supported by work on 4 enzyme families and 1 enzyme superfamily that found within each family and within the superfamily the special charge distribution is maintained [17].  Moreover, conserved electrostatic interaction networks have been demonstrated among TIM barrel proteins [18].

One rationale for this relationship is that eliminating a critical electrostatic perturbation can interfere with the proper binding and/or catalysis of ligands, since this could alter the electrostatic environment of the catalytic site in a way that no longer supported either function.  Support for this relationship between altered electrostatics and altered functionality is found in that among the 11 drug resistance mutations found in ionizable residues of the HIV reverse transcriptase 8 of them are in residues that exhibit electrostatic perturbations [4].  Another possibility is that it has been demonstrated that pKa perturbations tend to be more prevalent among buried residues and that buried residues tend to exhibit a higher degree of conservation since the ability of a protein to accommodate a mutation in a buried residue is often highly constrained by residue packing [19,20].  Given the surface accessibility of the DEAD box motif residues, the conserved electrostatics are more likely the result of the need to maintain catalytic functionality over evolutionary time.

Moreover, the results obtained in this study also provide important insights into the functionality of DEAD-box proteins.  It has been observed in numerous DEAD-box proteins that the binding of ligands, such as DNA or RNA, to the protein can exhibit an effect on the rate of ATP hydrolysis.  Furthermore, it has also been suggested that the binding of ligands might lead to conformational changes within the protein that could impact the residues of the catalytic site [1,2].  This study demonstrates that the electrostatic interaction network of the DEAD box glutamate has a greater number of interactions in the pre-catalytic state than in the non-catalytic state. Mutational analysis of several of these interacting residues demonstrated that the interactions are important for the maintenance of pre-catalytic electrostatics.  Thus it is hypothesized that the conformational changes brought about by ligand binding are critical to altering the electrostatic interaction network of the catalytic glutamate and in turn help to facilitate catalysis by introducing a pKa perturbation in the residue.

In all, the findings of this study as well as others that have been referenced throughout this discussion are growing increasingly suggestive that the conservation of protein physical states may be as critical to protein evolution and function as the more commonly used measure of residue conservation.